\documentclass[12pt]{article}
\usepackage{graphicx,psfrag,epsf}
\usepackage{enumerate}
\usepackage{natbib}
\usepackage{url} 
\usepackage{graphicx}
\usepackage{amssymb}
\usepackage{amsfonts}
\usepackage{epsfig}
\usepackage{mathrsfs}
\usepackage{amsmath,amsthm,amssymb,amscd}
\usepackage{MnSymbol}
\usepackage{rotating}
\usepackage{multirow}
\usepackage{float}
\usepackage[english]{babel}
\usepackage{floatpag}
\usepackage{enumerate}
\usepackage{threeparttable}

\newcommand\appendix@section[1]{\refstepcounter{section}\orig@section*{#1}\addcontentsline{toc}{section}{#1}}
\let\orig@section\section\g@addto@macro\appendix{\let\section\appendix@section}
\makeatother \makeatletter
\renewcommand\footnoterule{\kern-3\p@ \hrule width 1\columnwidth \kern 2.6\p@}
\newskip\@footindent
\renewcommand\@footindent{0pt}
\@addtoreset{footnote}{page}
\long\def\@makefntext#1{\@setpar{\@@par\@tempdima \hsize
\advance\@tempdima-\@footindent \parshape \@ne \@footindent
\@tempdima}\par \noindent \hbox to
\z@{\hss\@thefnmark£º\hspace{0.2em}}#1}

\def\@makefnmark{\hbox{\textsuperscript{\@thefnmark}}}
\makeatother

\newcommand{\sdn}{\sum_{i=1}^{n}\sum_{j=1}^{n}}
\newcommand{\sdtn}{\sum_{i=1}^{\tn}\sum_{j=1}^{\tn}}

\newcommand{\tz}{\tilde{Z}}
\newcommand{\ty}{\tilde{Y}}
\newcommand{\tx}{\tilde{X}}
\newcommand{\tl}{\tilde{L}}
\newcommand{\tr}{\tilde{R}}
\newcommand{\tn}{\tilde{n}}

\newcommand{\tvar}{\tilde{\varepsilon}}
\newcommand{\te}{\tilde{e}}
\newcommand{\tu}{\tilde{U}}
\newcommand{\tg}{\tilde{G}}
\newcommand{\bg}{\bar{G}}
\newcommand{\hbeta}{\hat{\beta}}
\newcommand{\tfw}{\tilde{\mathsf{w}}}
\newcommand{\fw}{\mathsf{w}}
\newcommand{\tps}{\tilde{\psi}}

\newtheorem{thm}{Theorem}[]
\newtheorem{lem}{Lemma}[]

\title{Regression analysis of doubly truncated data}
\author{  \vspace{2mm} Zhiliang Ying$^\ast$\\
 Department of Statistics \\
 Columbia University\\
 \vspace{5mm}
 {\small $^\ast$Corresponding author: \ zying@stat.columbia.edu}\\
 \vspace{2mm} Wen Yu\\
 Department of Statistics\\
 School of Management\\
 \vspace{5mm}
 Fudan Uninversity\\
 \vspace{2mm} Ziqiang Zhao  \\
 \vspace{5mm}
 Novartis Pharmaceuticals\\
 \vspace{2mm} Ming Zheng \\
 Department of Statistics\\
 School of Management\\
 Fudan Uninversity\\
 }
\date{}
\begin{document}
\maketitle
\newpage
\begin{abstract}
Doubly truncated data are found in astronomy, econometrics and survival analysis literature. They arise when each observation is confined to an interval, i.e., only those which fall within their respective intervals are observed along with the intervals. Unlike the more widely studied one-sided truncation that can be handled effectively by the counting process-based approach, doubly truncated data are much more difficult to handle. In their analysis of an astronomical data set, Efron and Petrosian (1999) proposed some nonparametric methods, including a generalization of Kendall's tau test, for doubly truncated data. Motivated by their approach, as well as by the work of Bhattacharya et al. (1983) for right truncated data, we proposed a general method for estimating the regression parameter when the dependent variable is subject to the double truncation. It extends the Mann-Whitney-type rank estimator and can be computed easily by existing software packages. We show that the resulting estimator is consistent and asymptotically normal. A resampling scheme is proposed with large sample justification for approximating the limiting distribution. The quasar data in Efron and Petrosian (1999) are re-analyzed by the new method. Simulation results show that the proposed method works well. Extension to weighted rank estimation are also given.
\end{abstract}

\vfill \hrule \vskip 6pt \noindent {\em MSC:}  \  

\newpage

\section{Introduction}
\label{sec1}

In their analysis of quasar data, Efron and Petrosian (1999) proposed nonparametric methods for doubly truncated data. Their methods deal with two common statistical issues: 1. testing independence between the explanatory variable and the dependent variable when the latter is subject to the double truncation; 2. estimating nonparametrically the marginal distribution of the response variable when the independence is true. For the first issue, they constructed an extension of Kendall's tau that corrects for possible bias due to the truncation. For the second issue, they applied the nonparametric EM algorithm to obtain a self-consistent estimator.

The existing literature contains many nonparametric methods for dealing with truncated data. Turnbull (1976) developed a general algorithm for finding the nonparametric maximum likelihood estimator of distribution for arbitrarily grouped, censored and truncated data. This estimator was obtained earlier by Lynden-Bell (1971) for singly truncated data. The large sample properties of Lynden-Bell's estimator were established by Woodroofe (1985). Wang, Jewell, and Tsai (1986), Keiding and Gill (1990) and Lai and Ying (1991a) applied the counting process-martingale techniques.

There is a substantial literature on regression analysis with the response variable subject to right or left truncation. Motivated from an application in astronomy, Bhattacharya, Chernoff, and Yang (1983) formulated the relationship between luminosity and red shift as a linear regression model in which the response variable is subject to right truncation. They extended the Mann-Whitney estimating function with a modification to correct for possible bias due to the truncation, and showed that their estimator is consistent and asymptotically normal. Tsui, Jewell, and Wu (1988) developed an iterative bias adjustment technique to estimate the regression parameter in the linear regression model. Tsai (1990) made use of Kendall's tau to construct tests for independence between the response and the explanatory variables. Lai and Ying (1991b) constructed a semiparametrically efficient estimator using rank based estimating functions. For modeling and analysis of truncated data in the econometrics literature, see Amemiya (1985) and Greene (2012), and references therein. For general biased sampling that contains truncation as special cases, we refer to recent works of Kim, Lu, Sit and Ying (2013) and Liu, Ning, Qin and Shen (2016).

Compared with singly truncated data, dealing with doubly truncated data is technically more challenging. Very few results have been obtained for doubly truncated data due to lack of explicit tools. Similar difficulties also arise for doubly censored data. Chang and Yang (1987) and Gu and Zhang (1993) discussed nonparametric estimators based on doubly censored data and established their asymptotic properties. Semiparametric regression M-estimators with doubly censored responses were studied by Ren and Gu (1997). For doubly truncated data, besides Efron and Petrosian (1999)'s work, Bilker and Wang (1996) extended the two-sample Mann-Whitney test, with parametric modeling of the truncation variables. Also for doubly truncated data, Shen (2013) considered semiparametric transformation models and used nonparametric EM algorithm as in Efron and Petrosian (1999) to obtain regression parameter estimation.

This paper proposes a general approach to estimating the regression parameter in the linear regression model when the response variable is subject to double truncation. An extended Mann-Whitney type loss function is introduced that takes into consideration of the double truncation. A Mann-Whitney-type rank estimator is then defined as its minimizer. The minimization can be carried out easily and efficiently using existing software packages. Additionally, a random perturbation approach is proposed for variance estimation and distributional approximation. By applying large sample theory for U-processes, a quadratic approximation is developed for the loss function and, as a consequence, the usual asymptotic properties are established for the proposed estimator. Large sample justification for the random perturbation approach is also given. Extensive simulation results are reported to assess the finite sample performance of the proposed method. The method is applied to the quasar data. Extensions to weighted Mann-Whitney-type pairwise comparisons that may improve efficiency are also proposed.

The rest of the paper is organized as follows. The next section introduces some basic notation and defines the doubly truncated linear regression which is the focus of this paper. In Section 3, we introduce an extension of the Mann-Whitney-type objective function for regression parameter estimation that adjusts for double truncation. The usual large sample properties of the proposed method are established in Section 4. In Section 5, we propose a weighting scheme for efficiency improvement. Sections 6 and 7 are devoted to simulation results and analysis of the quasar data, respectively. Some concluding remarks are given in Section 8. Some technical developments are given in the Appendix.

\section{Notation and model specification}
\label{sec2}

We will be concerned with the standard linear regression model
\begin{eqnarray}
\ty=\beta^{\top}\tx+\tvar,\label{lr}
\end{eqnarray}
where $\ty$ is the response variable, $\tx$ the $p$-dimensional covariate vector with $\beta$ the corresponding regression parameter vector and  $\tvar$ the error term that is independent of covariates. This model becomes much more complicated when the response variable $\ty$ is subject to double truncation. Specifically, let $\tl$ and $\tr$ denote the left and right truncation variables. The response $\ty$, the truncation pair $(\tl,\tr)$ and covariates $\tx$ are observed if and only if $\tl<\ty<\tr$. Throughout this paper, we will make the usual independent truncation assumption: $\ty$ and $(\tl, \tr)$ are conditionally independent given $\tx$ or, equivalently, $\tvar$ is independent of $(\tx,\tl,\tr)$. We will use $f$ and $F$ to denote respectively the density and distribution functions of $\tvar$.

Let $\tz=(\ty,\tx^{\top},\tl,\tr)^{\top}$ and denote by $\tz_1,\ldots,\tz_{\tn}$ $\tn$ independent and identically distributed (i.i.d.) copies of $\tz$. Because of truncation, for each $i$, $\tz_i$ is observed if and only if $\tl_i<\ty_i<\tr_i$. Let $n=\#\{i: \tl_i<\ty_i<\tr_i\}$, the number of observations. Furthermore, let $Z_i=(Y_i,X_i^{\top},L_i,R_i)^{\top}$, $i=1,\dots, n$ be the observed $\tz_i$'s with $\varepsilon_i$ the corresponding error terms.

There are two approaches to formulate the truncation data. The first one, as being used here, is from the missing data viewpoint with $\tz_i$, $i=1, \dots, \tn$ as the complete data. The second
one is to directly model the observed data, i.e. to assume that $Z_i$, $i=1,\dots, n$ are i.i.d. observations with joint density
\begin{eqnarray}
\frac{f(Y_i-\beta^{\top}X_i)}{F(R_i-\beta^{\top}X_i)-F(L_i-\beta^{\top}X_i)}h(L_i, R_i, X_i), \quad L_i<Y_i<R_i,\label{2.2}
\end{eqnarray}
where $h$ is the joint density of $(L_i, R_i, X_i^{\top})^{\top}$. It can be shown that these two approaches are essentially equivalent. We used the first approach in the next section to motivate our estimator. However, rigorous asymptotic properties will be developed based on the second formulation.

The following notation will be used. For each $i=1,\ldots,n$, let $L_i(\beta)=L_i-\beta^{\top}X_i$, $R_i(\beta)=R_i-\beta^{\top}X_i$ and $e_i(\beta)=Y_i-\beta^{\top}X_i$. Correspondingly, let $\tl_i(\beta)=\tl_i-\beta^{\top}\tx_i$, $\tr_i(\beta)=\tr_i-\beta^{\top}\tx_i$ and $\te_i(\beta)=\ty_i-\beta^{\top}\tx_i$, $i=1,\ldots,\tn$.

\section{Methods}
\label{sec3}

We are concerned with inference about the regression parameter $\beta$. If $\tz_1,\ldots,\tz_{\tn}$ were observed, one could use the following Mann-Whitney-type estimating equation (Jin, Ying, and Wei, 2001)
\begin{eqnarray}
\tu_{\tn}(\beta)=\sdtn(\tx_i-\tx_j)\mbox{sgn}\left\{\te_i(\beta)-\te_j(\beta)\right\}=0,\label{3.1}
\end{eqnarray}
where $\mbox{sgn}\{\cdot\}$ is the sign function. This estimating function is unbiased since, by symmetry, $\mathsf{E}(\mbox{sgn}\{\te_i(\beta)-\te_j(\beta)\}|\tx_i,\tx_j)=0$ when $\beta$ takes the true value. Under the double truncation, only those $\te_i(\beta)$ satisfying $\tl_i(\beta)<\te_i(\beta)<\tr_i(\beta)$ are observed. $\tu_{\tn}(\beta)$ would be biased if the summation on the right-hand-side of (\ref{3.1}) only include those observed pairs. However, this bias can be corrected if we impose an artificial symmetrical truncation with further restriction $\tl_j(\beta)<\te_i(\beta)<\tr_j(\beta)$. To this end, we define
\begin{eqnarray*}
U_n(\beta)&=&\sdtn I\left\{\tl_i(\beta)\vee\tl_j(\beta)<\te_i(\beta)<\tr_i(\beta)\wedge\tr_j(\beta),\tl_i(\beta)\vee\tl_j(\beta)<\te_j(\beta)<\tr_i(\beta)\wedge\tr_j(\beta)\right\} \nonumber\\
& &\times(\tx_i-\tx_j)\mbox{sgn}\left\{\te_i(\beta)-\te_j(\beta)\right\},
\end{eqnarray*}
where $I\{\cdot\}$ is the indicator function and $\wedge$ ($\vee$) is the minimum (maximum) operator. Again, by symmetry, $U_n(\beta)$ is an unbiased estimating function as its conditional expectation given the $\tl_i,\tr_i,\tx_i$ is zero. Furthermore, the non-zero terms in $U_n(\beta)$ are observed because of the constraints imposed. In fact, we can write
\begin{eqnarray*}
U_n(\beta)=\sdn I\left\{L_j(\beta)<e_i(\beta)<R_j(\beta),L_i(\beta)<e_j(\beta)<R_i(\beta)\right\}(X_i-X_j)\mbox{sgn}\left\{e_i(\beta)-e_j(\beta)\right\}.
\end{eqnarray*}

Estimating function $U_n(\beta)$ is a step function, thus discontinuous. Finding root of a discontinuous function is typically not easy, especially for multidimensional cases. However, in the case of no truncation, finding root of $\tu_{\tn}(\beta)$ is equivalent to minimizing an $L_1$-type loss function $\tg_{\tn}(\beta)=\sdtn|\te_i(\beta)-\te_j(\beta)|=\sdtn|\ty_i-\ty_j-\beta^\top(\tx_i-\tx_j)|$, which is convex (Jin et al., 2001). In fact, this is a linear programming problem (Koenker and Bassett, 1978).

For doubly truncated data, we propose the following loss function
\begin{eqnarray}
{G}_n(\beta)=\sdn\left|\left[\left(e_i(\beta)-e_j(\beta)\right)\wedge(R_j-Y_j)\wedge(Y_i-L_i)\right]\vee(L_j-Y_j)\vee(Y_i-R_i)\right|.\label{3.4}
\end{eqnarray}
Clearly, $G_n(\beta)$ becomes $\tg_{\tn}(\beta)$ when there is no truncation, i.e. $\tl_i\equiv-\infty$ and $\tr_i\equiv\infty$. Unlike $\tg_n(\beta)$, $G_n(\beta)$ is generally not a convex function. To see this, let $\underline{D}_{ij}=(L_j-Y_j)\vee(Y_i-R_i)$, $\overline{D}_{ij}=(R_j-Y_j)\wedge(Y_i-L_i)$, $Y_{ij}=Y_i-Y_j$ and $X_{ij}=X_i-X_j$. We have
\begin{eqnarray*}
G_n(\beta)=\sdn\left|\left(Y_{ij}-\beta^\top X_{ij}\right)\wedge\overline{D}_{ij}\vee\underline{D}_{ij}\right|.
\end{eqnarray*}
Since for any constants $a<b$, function $g(x)=|x\wedge a\vee b|$ is neither convex nor concave, $G_n(\beta)$ is generally not a convex function.

To see that minimizing the loss function $G_n(\beta)$ induces a consistent estimator, let
\begin{eqnarray}
\bg(\beta)=\mathsf{E}\left\{\left|\left[\left(e_i(\beta)-e_j(\beta)\right)\wedge(R_j-Y_j)\wedge(Y_i-L_i)\right]\vee(L_j-Y_j)\vee(Y_i-R_i)\right|\right\}.\label{3.5}
\end{eqnarray}
It can be proved that under mild conditions, $\bg(\beta)$ is the limit of $[n(n-1)]^{-1}G_n(\beta)$ uniformly for $\beta$ over a compact set. Differentiation of the right-hand-side of (\ref{3.5}) can be carried out by interchanging the differentiation and the expectation. Except on a set with zero probability, the derivative of the term inside the expectation sign is equal to
\begin{eqnarray}
I\left\{(L_j-Y_j)\vee(Y_i-R_i)<e_i(\beta)-e_j(\beta)<(R_j-Y_j)\wedge(Y_i-L_i)\right\}(X_i-X_j)\mbox{sgn}\left\{e_i(\beta)-e_j(\beta)\right\}.\label{3.6}
\end{eqnarray}
From Lemma \ref{lem1} in the Appendix, we can see that
\begin{eqnarray*}
(L_j-Y_j)\vee(Y_i-R_i)<e_i(\beta)-e_j(\beta)<(R_j-Y_j)\wedge(Y_i-L_i)
\end{eqnarray*}
occurs if and only if $L_j(\beta)<e_i(\beta)<R_j(\beta)$ and $L_i(\beta)<e_j(\beta)<R_i(\beta)$. Thus, by symmetry, the expectation of (\ref{3.6}) equals to zero when $\beta$ takes its true value, implying that $\bg(\beta)$ has a minimizer at the true value of $\beta$.

Although $G_n(\beta)$ is generally not convex, in many cases it has a global minimizer, especially when the truncation is mild, making $G_n(\beta)$ close to $\tg_n(\beta)$. In our experience, we find that optimization functions in standard software packages can be used effectively to find the minimizer of $G_n(\beta)$ directly. For instance, `{\it fminsearch}' in the `Optimization Toolbox' of MATLAB may be used for finding the global minimizer.

Alternatively, the computation can be formulated as an iterative $L_1$-minimization problem. To be specific, consider the following modification of (\ref{3.4})
\begin{eqnarray*}
G^{(m)}_n(\beta,b)=\sdn I\left\{L_j(b)<e_i(b)<R_j(b),L_i(b)<e_j(b)<R_i(b)\right\}\left|e_i(\beta)-e_j(\beta)\right|.
\end{eqnarray*}
Let $\hat{\beta}_{(0)}$ be an initial estimate, for instance, the estimate of $\beta$ by ignoring double truncation. An iterative algorithm is given by
\begin{eqnarray*}
\hat{\beta}_{(k)}=\mbox{arg}\min_\beta G^{(m)}_n(\beta,\hat{\beta}_{(k-1)}) \ \ (k\geqslant1).
\end{eqnarray*}
Note that in each iteration, $G^{(m)}(\beta,\hat{\beta}_{(k-1)})$ is an $L_1$-type objective function, and $\hat{\beta}_{(k)}$ solves the equation
\begin{eqnarray*}
\sdn& &I\left\{L_j(\hat{\beta}_{(k-1)})<e_i(\hat{\beta}_{(k-1)})<R_j(\hat{\beta}_{(k-1)}),
L_i(\hat{\beta}_{(k-1)})<e_j(\hat{\beta}_{(k-1)})<R_i(\hat{\beta}_{(k-1)})\right\}\\
& &\times(X_i-X_j)\mbox{sgn}\left\{e_i(\beta)-e_j(\beta)\right\}=0,
\end{eqnarray*}
If $\hat{\beta}_{(k)}$ converges to a limit as the number of $k\to\infty$, then the limit must satisfy $U_n(\beta)=0$.

Let $\hbeta_n$ denote the minimizer of $G_n(\beta)$ over a suitable parameter space. We show in Section 4 that $\hbeta_n$ is consistent and asymptotically normal under suitable regularity conditions. Like most estimators derived from non-smooth objective functions or discontinuous estimating functions, there is no simple plug-in variance estimator. Following Jin et al. (2001), we propose using resampling approach based on random weighting. Specifically, generate i.i.d. nonnegative random variables $W_i$, $i=1,\ldots,n$, with mean $\mu$ and variance $4\mu^2$. Define the following perturbed version of $G_n(\beta)$
\begin{eqnarray}\label{3.7}
G^\ast_n(\beta)=\sdn\left(W_i+W_j\right)\left|\left[\left(e_i(\beta)-e_j(\beta)\right)\wedge(R_j-Y_j)\wedge(Y_i-L_i)\right]\vee(L_j-Y_j)\vee(Y_i-R_i)\right|
\end{eqnarray}
and let $\hbeta^\ast=\mbox{argmin}_\beta G^\ast_n(\beta)$. We show in Section 4 that the conditional distribution of $\sqrt{n}(\hbeta^\ast-\hbeta_n)$ given data converges to the same limiting distribution as that of $\sqrt{n}(\hbeta_n-\beta_0)$, where $\beta_0$ is the true value of $\beta$. By repeatedly generating $\{W_i, i=1,\ldots,n\}$, we can obtain a large number of replications of $\hbeta^\ast$. Then the conditional distribution of $\sqrt{n}(\hbeta^\ast-\hbeta_n)$ given data can be approximated arbitrarily closely.

\section{Large sample theory}
\label{sec4}

This section is devoted to the development of a large sample theory for the methods proposed in the preceding section. Assume that $Z_i$, $i=1,\ldots,n$ are i.i.d. observations from (\ref{2.2}). Let $\beta_0$ denote the true parameter value. As we mention in Section 3, $\hbeta_n$ is the minimizer of $G_n(\beta)$ over a parameter space $B$. We shall assume that $B$ is compact and $\beta_0$ is an interior point of $B$.
Let
\begin{eqnarray*}
\xi(Z_i,Z_j,\beta)=I\left\{L_j(\beta)<e_i(\beta)<R_j(\beta),L_i(\beta)<e_j(\beta)<R_i(\beta)\right\}(X_i-X_j)\mbox{sgn}\left\{e_i(\beta)-e_j(\beta)\right\}
\end{eqnarray*}
and $V=\mathsf{E}[\xi(Z_i,Z_j,\beta_0)\xi^\top(Z_i,Z_k,\beta_0)]$. Also, let $A=\partial^2\bg/\partial\beta\partial\beta^\top|_{\beta=\beta_0}$. The following regularity conditions will be used.

\begin{enumerate}
\item [A1] The error density $f$ is bounded and has a bounded and continuous derivative.
\item [A2] The covariate vector has a bounded second moment, i.e., $\mathsf{E}(\|X\|^2)<\infty$.
\item [A3] The true parameter value $\beta_0$ is the unique global minimizer of the limiting loss function $\bg(\beta)$ over $B$.
\item [A4] The second derivative of $\bg(\beta)$ at $\beta_0$ is nonsingular, i.e., $A$ strictly positive definite.
\end{enumerate}

Conditions A1, A2 and A4 are mild conditions. Condition A3 is generally not verifiable. It is assumed to guarantee that the proposed estimator is consistent. The following theorem gives out the asymptotic properties of the proposed estimator.

\begin{thm}\label{thm1}
Under conditions A.1-A.4, $\hbeta_n$ is consistent and $\sqrt{n}(\hbeta_n-\beta_0)$ converges in distribution to $N(0,A^{-1}VA^{-1})$.
\end{thm}

The objective function $G_n(\cdot)$ is a typical $U$-process of order 2. Thus, we can apply results on quadratic approximations $U$-processes to prove the above result. The details are provided in the Appendix.

The limiting covariance matrix is, among other things, a functional of the error density. Thus, direct variance estimation involves density estimation. In principle, one may apply the nonparametric method proposed by Efron and Petrosian (1999) to the residuals to first estimate the error distribution and then, via smoothing, density. As being proposed in Section 3, we approach the variance estimation through random weighting. The theoretical justification of this approach is given by the following theorem. The proof of the theorem is given in the Appendix .

\begin{thm}\label{thm2}
Let $\hbeta^\ast$ be the minimizer of the perturbed loss function $G_n^\ast(\beta)$ as defined by (\ref{3.7}). Then under conditions A.1-A.4, the conditional distribution of $\sqrt{n}(\hbeta^\ast-\hbeta_n)$ given $Z_1,\ldots,Z_n$ converges in probability to $N(0,A^{-1}VA^{-1})$. In particular, the conditional covariance matrix of $\hbeta^\ast$ given $Z_1,\ldots,Z_n$ converges to $A^{-1}VA^{-1}$.
\end{thm}

\section{Weighted estimation}
\label{sec5}

It is well known that choosing proper weights can improve the estimating efficiency of the rank estimator; see, for example, Hajek and Sidak (1967), Prentice (1978), Harrington and Fleming (1982) and Jin et al. (2003). For the full data, we may extend the estimating function $\tu_{\tn}(\beta)$ in (\ref{3.1}) by assigning weights to its summands. Specifically, we consider the following weighted estimating function
\begin{eqnarray}
\tu_{\tn,\tfw}(\beta)=\sdn\tfw_{ij}(\tx_i-\tx_j)\mbox{sgn}\left\{\te_i(\beta)-\te_j(\beta)\right\},\label{5.1}
\end{eqnarray}
where the weights $\tfw_{ij}$, which may depend on $\beta$, are symmetric, i.e., $\tfw_{ij}=\tfw_{ji}$. By symmetry, we can easily see that the estimating function is unbiased, i.e., $\mathsf{E}[\tu_{\tn,\tfw}(\beta_0)]=0$. The choice of $\tfw_{ij}\equiv1$ corresponds to the Wilcoxon-Mann-Whitney statistic. It is asymptotically efficient when $\tvar$ in model (\ref{lr}) follows the standard logistic distribution. Under this weighting scheme, $\tu_{\tn,\tfw}(\beta)$ reduces to the unweighted estimating function $\tu_{\tn}(\beta)$. Another commonly used weighting scheme in rank estimation is that of the log-rank, which is asymptotically efficient when $\tvar$ follows the extreme minimum value distribution. Let $\tfw_{ij}=\tfw_{ij}(\beta)=\tps_{\tn}(\beta,\te_i(\beta)\wedge\te_j(\beta))$, where $\tps_{\tn}(b,t)=(\sum_{i=1}^{\tn}I\{\te_i(b)\geqslant t\})^{-1}$. We show in Lemma \ref{lem2} in the Appendix that such choice of $\tfw_{ij}$ leads $\tu_{\tn,\tfw}(\beta)$ to become the log-rank estimation function for $\beta$.

For the doubly truncated data, similar to (\ref{5.1}), we can also introduce weights to the proposed estimating function $U_n(\beta)$, that is, to consider
\begin{eqnarray*}
U_{n,\fw}(\beta)=\sdn\fw_{ij}(\beta)I\left\{L_j(\beta)<e_i(\beta)<R_j(\beta),L_i(\beta)<e_j(\beta)<R_i(\beta)\right\}
(X_i-X_j)\mbox{sgn}\left\{e_i(\beta)-e_j(\beta)\right\},
\end{eqnarray*}
where the $\fw_{ij}$ are again symmetric, i.e. $\fw_{ij}=\fw_{ji}$. Mimicking the full-data situation, we treat $\fw_{ij}=1$ as the Wilcoxon weight, corresponding to the originally proposed estimating function $U_n(\beta)$. For the log-rank version, we let $\fw_{ij}(\beta)=\psi_n(\beta,e_i(\beta)\wedge e_j(\beta))$, where $\psi_n(b,t)=(\sum_{i=1}^nI\{e_i(b)\geqslant t\})^{-1}$. Other weighting schemes can also be considered. Though the data is subject to double truncation, we still expect, as simulation results in the subsequent section also indicate, that proper choices of weights will generally improve the estimation efficiency.

Similar to $U_n(\beta)$, $U_{n,\fw}(\beta)$ is discontinuous and solving $U_{n,\fw}(\beta)=0$ directly may not be easy. As in the case of the log-rank estimation function, $\fw_{ij}$ typically depends on $\beta$. Write $\fw_{ij}=\fw_{ij}(\beta)$. We consider loss function
\begin{eqnarray*}
G_{n,\fw}(\beta,b)=\sdn\fw_{ij}(b)\left|\left[\left(e_i(\beta)-e_j(\beta)\right)\wedge(R_j-Y_j)\wedge(Y_i-L_i)\right]\vee(L_j-Y_j)\vee(Y_i-R_i)\right|.
\end{eqnarray*}
By differentiating with respect to $\beta$, it is easily seen that
\begin{eqnarray}\label{5.2}
\frac{\partial G_{n,\fw}(\beta,b)}{\partial\beta}\Big|_{b=\beta}&=&\sdn\fw_{ij}(\beta)I\left\{L_j(\beta)<e_i(\beta)<R_j(\beta),L_i(\beta)<e_j(\beta)<R_i(\beta)\right\}\nonumber\\
& &\times(X_i-X_j)\mbox{sgn}\left\{e_i(\beta)-e_j(\beta)\right\},
\end{eqnarray}
which becomes the weighted estimating function $U_{n,\fw}(\beta)$. Therefore, we propose the following iterative algorithm. First set the initial $b$ to be $\hbeta_{(0)}^{\fw}$, and then find the estimator iteratively through $\hbeta_{(k)}^{\fw}=\mbox{argmin}_\beta G_{n,\fw}(\beta,\hbeta_{(k-1)}^{\fw})$, $k\geqslant1$. From (\ref{5.2}) we see that if $\hbeta_{(k)}^{\fw}$ converges to a limit, say $\hbeta_{n}^{\fw}$, as $k$ goes to infinity, then the limit satisfies $U_{n,\fw}(\hbeta_{n}^{\fw})=0$.

For the weights $\fw_{ij}$ with form $\psi_n(\beta,e_i(\beta)\wedge e_j(\beta))$, where $\psi_n(b,t)$ may depend on the data, we assume the following condition.

\begin{enumerate}
\item [A5] There exists a deterministic function $\psi(t)$ such that $\sup_t|\psi_n(\beta_0,t)-\psi(t)|=o_p(n^{-\eta})$ for some $\eta>0$.
\end{enumerate}

The asymptotic properties of the weighted estimator is given by the following theorem.

\begin{thm}\label{thm3}
Under conditions A.1-A.5, $\hbeta_{n}^{\fw}$ is consistent and $\sqrt{n}(\hbeta_{n}^{\fw}-\beta_0)$ converges in distribution to $N(0,A_{\fw}^{-1}V_{\fw}A_{\fw}^{-1})$.
\end{thm}

Matrices $A_{\fw}$ and $V_{\fw}$ are the asymptotic slope and the covariance matrices for the weighted estimating function $U_{n,\fw}$ that reduce to $A$ and $V$ when $\fw_{ij}=1$. As noted in Jin et al. (2003), when using the above algorithm, for a fixed $k$, $\hbeta_{(k)}^{\fw}$ is itself a legitimate estimator, i.e. it is consistent and asymptotically normal. Specifically, we have the following result.

\begin{thm}\label{thm4}
Under conditions A.1-A.5, for each $k\geqslant0$, $\sqrt{n}(\hbeta_{(k)}^{\fw}-\beta_0)$ converges in distribution to a normal distribution with zero mean and some variance-covariance matrix.
\end{thm}

In view of the above result, one may in practice consider the proposed iterative algorithm only for a relatively small number of the iterations to obtain a reasonable estimator. In our simulation study, we set the number of iterations to be 3 to get the log-rank estimate. We also iterated the algorithm until the difference between successive estimates attains a pre-specified accuracy as ``convergence''. We found that $\hbeta_{(k)}^\fw$ converged in all the cases and the converged estimate was quite close to the $\hbeta_{(k)}^\fw$ after 3 iterations.

For the variance estimation, we may follow Jin et al. (2003) by applying the random weighting approach. We introduce the following perturbed version of $G_{n,\fw}(\beta,b)$:
\begin{eqnarray*}
G_{n,\fw}^{\ast}(\beta,b)=\sdn(W_i+W_j)\fw_{ij}(b)\left|\left[\left(e_i(\beta)-e_j(\beta)\right)\wedge(R_j-Y_j)\wedge(Y_i-L_i)\right]\vee(L_j-Y_j)\vee(Y_i-R_i)\right|,
\end{eqnarray*}
where $W_i$, $i=1,\ldots,n$, are i.i.d. nonnegative random variables with mean $\mu$ and variance $4\mu^2$. The perturbed estimate is solved by exactly following the above iterative algorithm. We first obtain $\hbeta^\ast$ from minimizing $G_{n,\fw}^\ast$ by setting $\fw_{ij}(b)=1$. Note that this $\hbeta^\ast$ is just the minimizer of (\ref{3.7}). Then let $\hbeta^\ast_{(0)}=\hbeta^\ast$, and iterate the value of the estimate by $\hbeta^\ast_{(k)}=\mbox{argmin}_\beta G_{n,\fw}^{\ast}(\beta,\hbeta^\ast_{(k-1)})$. It is important to point out that here the number of iteration should stay the same as that for solving the point estimate. The asymptotic distribution of $\sqrt{n}(\hbeta_{(k)}^{\fw}-\beta_0)$ can be approximated by the conditional distribution of $\sqrt{n}({\hbeta_{(k)}}^\ast-\hbeta_{(k)}^{\fw})$ given the observed data. By repeatedly generating the $W_i$ sequences, we can obtain many realizations of $\hbeta^\ast_{(k)}$ and make inference based on the empirical distribution of the realized $\hbeta^\ast_{(k)}$'s.

\section{Simulation study}
\label{sec6}

In this section, simulation studies were conducted to assess the finite sample performance of the proposed method. For model (\ref{lr}), we considered a two-dimensional covariate vector, i.e., $\tx=(\tx_1,\tx_2)^\top$, where $\tx_1$ and $\tx_2$ were independently drawn from a binomial distribution with success probability $0.5$ and uniform distribution on $[0,2]$, respectively. We set the two regression coefficients, denoted by $\beta_1$ and $\beta_2$, to be 0 and 1. For the error distribution $F$, three distributions, standard normal distribution, standard logistic distribution and extreme minimum value (EV) distribution, were used. We considered two truncation schemes. The first one was covariate-independent, with the truncation variables $\tl$ and $\tr$ being independently generated from uniform distribution on $[c_1, 1]$ and uniform distribution on $[1,c_2]$, respectively. The second one was covariate-dependent, with $\tl$ and $\tr$ being independently generated from uniform distribution on $[c_3, \tx_1+\tx_2/2]$ and uniform distribution on $[\tx_1+\tx_2/2,c_4]$. The constants $c_1$ to $c_4$ were chosen to yield about $30\%$ percentage of truncation under various error distributions (with both left and right truncation proportions being of $15\%$). The observable sample size $n$ was chosen to be $200$, $300$ and $400$. Under each scenario, $1,000$ replications were carried out. We first used the originally proposed loss function (\ref{3.4}), which corresponds to the Wilcoxon weight in the view of the weighted approach, to get the estimate. Then we considered the log-rank weight, using the proposed iterative algorithm with the iteration number being 3, as we mention in Section 5. The minimization was implemented using the MATLAB function `{\it fminsearch}' in the `Optimization Toolbox' of MATLAB, which uses a simplex search method to find the minimizer. For estimating standard errors using the proposed resampling approach, 500 sets of i.i.d. random variables $W_i$, $i=1,\ldots,n$, of Gamma(0.25,0.5) were generated.

Besides the proposed estimates, we also calculated ``naive" estimates for the regression coefficients by ignoring the truncation. That is, we treated the observed data as data without double truncation, and solved the Mann-Whitney type estimating equation (\ref{3.1}) for the estimates. The random weighting approach proposed by Jin et al. (2001) was applied to get the estimated standard errors. For all the estimates, we recorded the average bias, empirical standard error, the average of the standard errors estimated from the random weighting approach, and the empirical coverage probability of the $95\%$ Wald-type confidence intervals. The results under covariate-independent truncation scenario are summarized in Table 1, while the results under covariate-dependent truncation are in Table 2.

\begin{center}
[Insert Table 1 here]
\end{center}

\begin{center}
[Insert Table 2 here]
\end{center}

We found that under covariate-independent truncation, the naive estimate for $\beta_1$ still had reasonable performance, but the naive estimate for $\beta_2$ was obviously biased, resulting in poor empirical coverage for the corresponding confidence interval. Under the covariate-dependent censoring, both naive estimates for $\beta_1$ and $\beta_2$ were biased and the empirical coverage probabilities of the confidence intervals were far less than the nominal level. However, under all scenarios, the proposed estimates obtained from the original loss function (i.e., Wilcoxon weight) and log-rank weight with $k=3$ were both essentially unbiased. The average of the standard error estimates were quite close to the corresponding empirical standard errors. The empirical coverage probabilities of the Wald-type confidence intervals were close to the nominal level. For the normally distributed random error, the estimates with the two weighting schemes had comparable efficiency. For the logistic random error, the Wilcoxon weight gave slightly more efficient estimates than those with the log-rank weight, while for the extreme minimum value random error, the estimate with log-rank weight was significantly more efficient. The results implied that for the doubly truncated data, one could still expect substantial efficiency improvement if a proper weighting scheme was chosen, as one would expect for the case with no truncation. In general, the simulation results showed that the proposed method worked well for practical sample sizes.

\begin{figure}
\begin{center}
\includegraphics[scale=0.28]{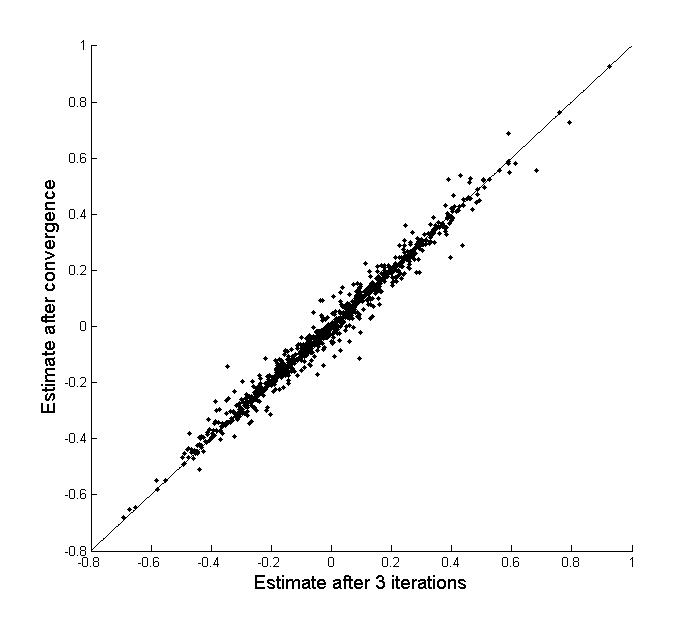}
\includegraphics[scale=0.28]{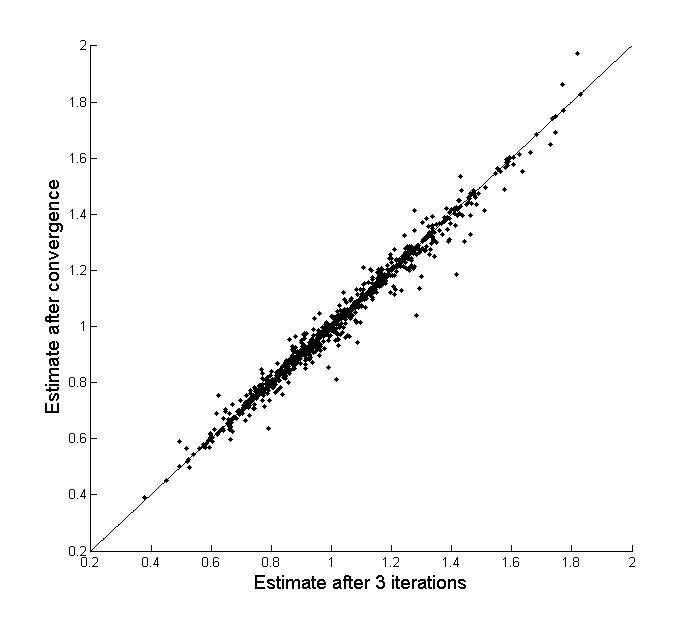}
\includegraphics[scale=0.28]{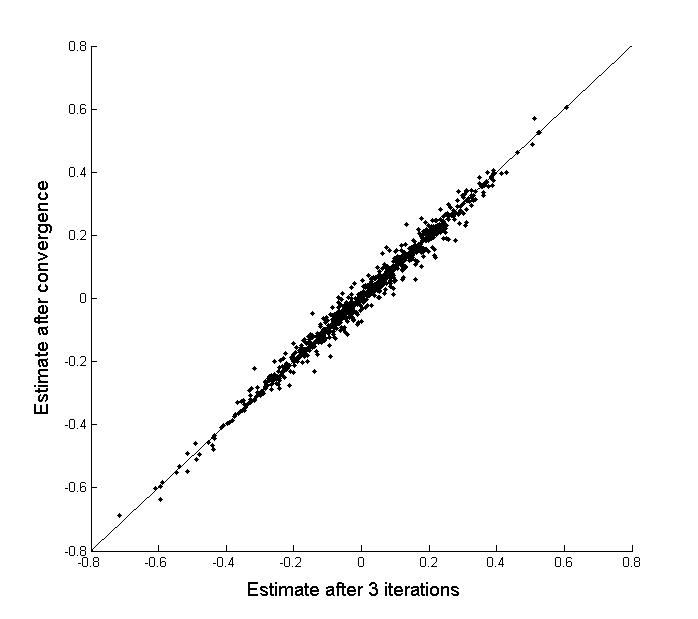}
\includegraphics[scale=0.28]{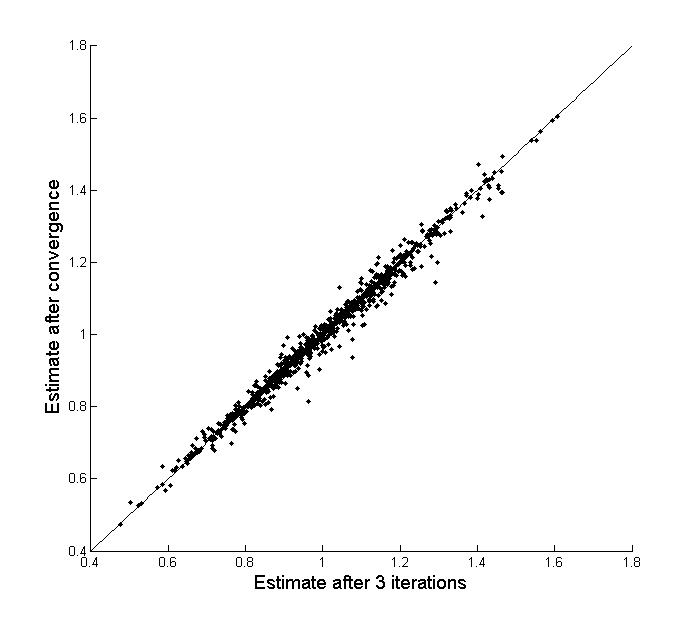}
\includegraphics[scale=0.28]{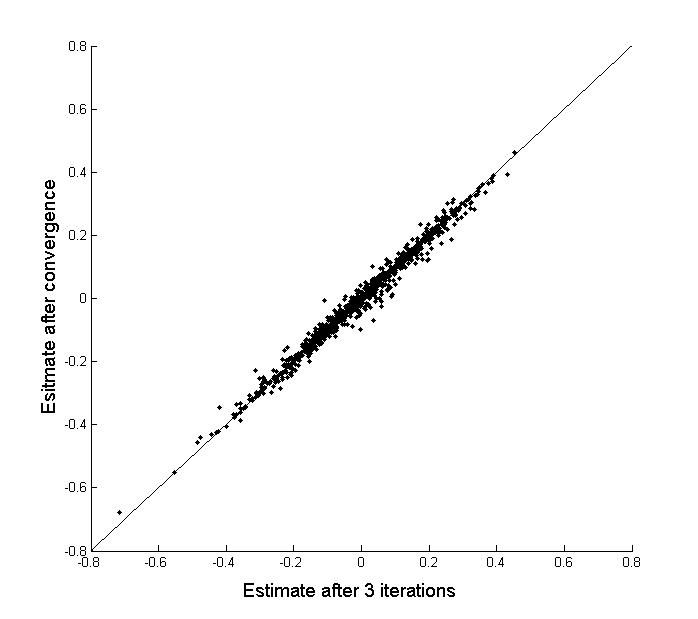}
\includegraphics[scale=0.28]{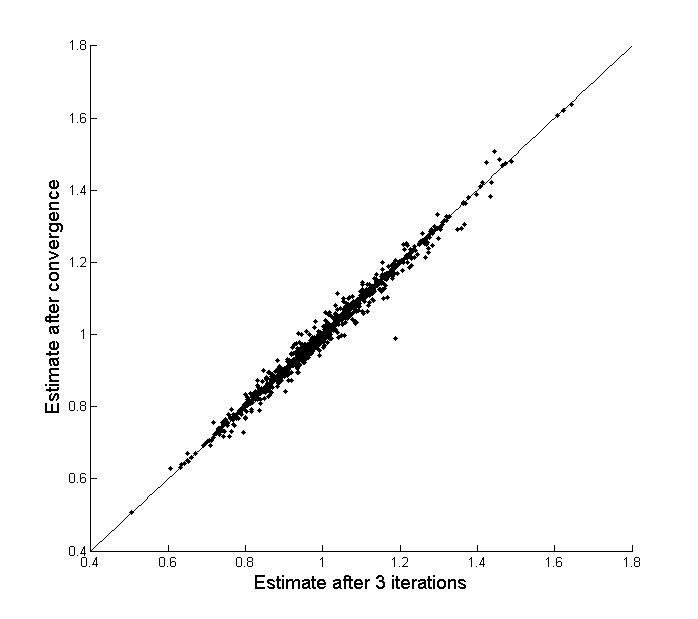}
\end{center}
\caption{Scatter plots of the estimates after 3 iterations against estimates after convergence. The error distribution was EV and the truncation was covariate-dependent. The top panel corresponds to $n=200$, the middle panel corresponds to $n=300$, and the bottom panel corresponds to under $n=400$. The left ones are for $\beta_1$ and the right ones are for $\beta_2$. \label{fig1}}
\end{figure}

We also examined the difference between the log-rank estimates with 3 iterations versus those obtained after convergence. The algorithm was treated as convergence in the sense that the sum of absolute component differences between two consecutive estimates was less than $0.01$. We took EV error distribution and covariate-dependent truncation for illustration. The estimates with 3 iterations and at convergence were plotted for the two regression parameters under different sample sizes. In Figure \ref{fig1}, the top panel corresponds to the plots for $\beta_1$ and $\beta_2$ under $n=200$, the middle panel corresponds to the plots for $\beta_1$ and $\beta_2$ under $n=300$, and the bottom panel corresponds to the plots for $\beta_1$ and $\beta_2$ under $n=400$. The two sets of estimates were quite similar, implying that a small number of iterations (such as 3) was sufficient. The situation was quite similar for the other error distributions and truncation mechanisms.

\begin{center}
[Insert Figure \ref{fig1} here]
\end{center}

\section{Application to quasar data}
\label{sec7}

We applied the proposed methods to the quasar data analyzed by Efron and Petrosian (1999). The dataset consists of quadruplets $(z_i,m_i,a_i,b_i)$, $i=1,\ldots,n$, where $z_i$ is the redshift of the $i$th quasar, $m_i$ is its apparent magnitude, and the two numbers $a_i$ and $b_i$ are lower and upper truncation bounds on apparent magnitude, respectively. Quasars with $m_i$ above $b_i$ were too dim to yield dependable redshifts, while the lower limit $a_i$ was used to avoid confusion with nonquasar steller objects. Thus, the apparent magnitude was doubly truncated. In this study $a_i=16.08$ remains the same for all $i$, and $b_i$ varies between 18.494 and 18.93. The full dataset has $n=1,052$ quasars.

Father quasars tend to have bigger values of $m_i$. According to Hubble's law, one can transform apparent magnitudes into a luminosity measurement which should be independent of distance. The transformation depends on the cosmological model supposed. Following the Einstein-deSitter cosmological model (Weinberg, 1972), one can obtain the log luminosity values $y_i$ from a formula
\begin{eqnarray}
y_i=t(z_i,m_i)=19.894-2.303\frac{m_i}{2.5}+\log(Z_i-Z_i^{\frac{1}{2}})-\frac{1}{2}\log(Z_i),\label{6.1}
\end{eqnarray}
where $Z_i=1+z_i$. Larger values of $y_i$ correspond to intrinsically brighter quasars. The truncation limits $L_i$ and $R_i$ for $y_i$ are obtained by applying (\ref{6.1}) to $a_i$ and $b_i$, i.e., $L_i=t(z_i, a_i)$ and $R_i=t(z_i, b_i)$.

The main purpose of the quasar investigation is to study luminosity evolution. Quasars may have been intrinsically brighter in the early universe and evolved toward a dimmer state as time went out. However, if there is no luminosity evolution, $y_i$ should be independent of $z_i$ except for truncation effects. Thus, testing the absence of luminosity evolution amounts to testing for independence. A convenient one-parameter model for luminosity evolution says that the expected log luminosity increases linearly as $\theta\log(1+z)$, with $\theta=0$ corresponding to no evolution. If $\theta$ is a hypothesized value of the evolution parameter, instead of directly testing for the independence of $y_i$ and $z_i$, Efron and Petrosian (1999) tested the null hypothesis that $\mathsf{H}_\theta$: $y_i(\theta)=y_i-\theta\log(1+z_i)$ is independent of $z_i$, using their proposed approach. Correspondingly, in their analysis, the truncation regions for $y_i(\theta)$ also changed with $\theta$, that is, $L_i(\theta)=L_i-\theta\log(1+z_i)$ and $R_i(\theta)=R_i-\theta\log(1+z_i)$.

\begin{figure}
\begin{center}
\includegraphics[scale=0.45]{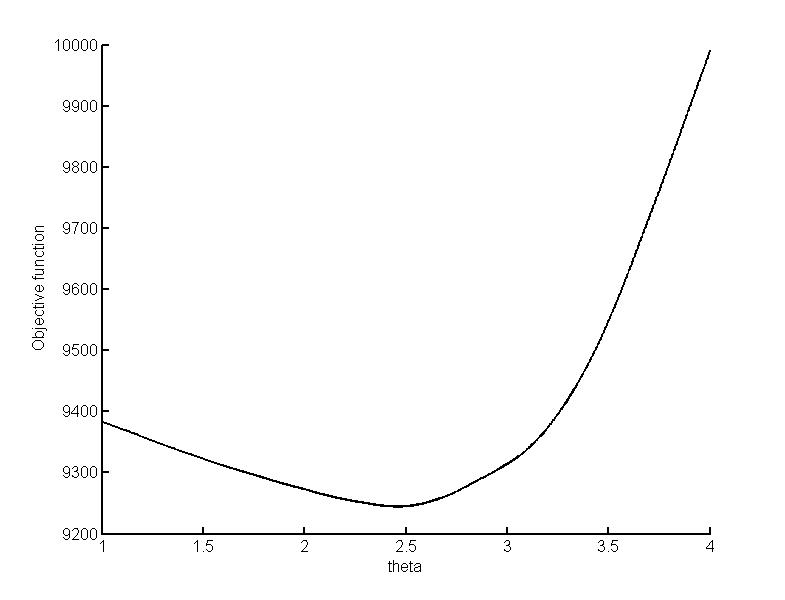}
\end{center}
\caption{Results from the quasar data analysis. The curve of the loss function $G_n(\theta)$ against $\theta$ within the range from $1$ to $4$. \label{fig2}}
\end{figure}

Since the one-parameter model for luminosity evolution assumes linear relationship between the expected log luminosity and $\log(1+z)$, it is quite natural to consider the following linear model
\begin{eqnarray}
y_i=\theta\log(1+z_i)+\varepsilon_i,\label{6.2}
\end{eqnarray}
where the response $y_i$ is subject to double truncation with the truncation region $[L_i,R_i]$, $\varepsilon_i$ is independent of $z_i$, and the evolution parameter $\theta$ becomes the unknown regression parameter. We can estimate $\theta$ by our proposed method. To make comparison, we used the same subset selected by Efron and Petrosian (1999) with $n=210$ to do the analysis. Here we considered the original loss function $G_n(\theta)$ defined in (\ref{3.4}). The point estimate, denoted by $\hat{\theta}_n$, was obtained by minimizing $G_n(\theta)$. Figure \ref{fig2} plots the curve $G_n(\theta)$ against $\theta$ within the range from $1$ to $4$.

\begin{center}
[Insert Figure \ref{fig2} here]
\end{center}

The estimate $\hat{\theta}_n$, which is the minimizer of the displayed loss function, was $2.458$. The proposed random weighting approach is used to estimate the standard error of $\hat{\theta}_n$. Five hundred draws of i.i.d. random variables following Gamma(0.25,0.5) were generated. The estimated standard error was $0.641$. Consequently, an approximate $90\%$ Wald-type confidence interval was $[1.40, 3.51]$. Under the linear model (\ref{6.2}), the hypothesis of no evolution, i.e.,  $\mathsf{H}_0$: $y_i$ is independent of $z_i$, is equivalent to $\mathsf{H}_0: \theta=0$. To test for $\mathsf{H}_0: \theta=0$ against a positive evolution parameter $\mathsf{H}_a: \theta>0$, a Wald-type test statistic can be used. The test statistic equaled to the ratio of $\hat{\theta}_n$ and its estimated standard error, giving the value of $3.835$. The corresponding one-sided $p$-value was about $6\times10^{-5}$, implying rejection of the null hypothesis of no evolution in favor of a positive value of $\theta$ at any commonly used significance level.

The tau test proposed by Efron and Petrosian (1999) for the no evolution hypothesis has an one-sided $p$-value $0.015$. At $0.05$ significance level, their test also rejected $\mathsf{H}_0$ in favor of a positive value of $\theta$, but failed to do so at $0.01$ significance level. By inverting their test statistic, Efron and Petrosian (1999) obtained a point estimate for $\theta$ with the value of $2.38$ and an approximate $90\%$ central confidence interval $[1.00, 3.20]$ which is slightly longer than the proposed Wald-type confidence interval.

The proposed approach is easy to handle multiple covariates. Here we further considered the following model with linear and quadratic term
\begin{eqnarray*}
y_i=\theta_1\log(1+z_i)+\theta_2\left[\log(1+z_i)\right]^2+\varepsilon_i,
\end{eqnarray*}
where $\varepsilon_i$ is independent of $z_i$ and $\theta_1$ and $\theta_2$ are unknown regression parameters. The regression parameters were estimated by minimizing (\ref{3.4}), and the standard errors were estimated by the random weighting method with 500 i.i.d. Gamma(0.25,0.5) random variables being generated. The corresponding $p$-values of significance test for $\mathsf{H}_0: \theta_j=0$ against $\mathsf{H}_a: \theta_j\ne0$, $j=1,2$, were calculated. The results are summarized in Table 3.

\begin{center}
[Insert Table 3 here]
\end{center}

The significance tests showed that the effect of linear term, $\theta_1$, was statistically significantly different from $0$, while that of the quadratic term, $\theta_2$, was apparently not. This provided some evidence to say the one-parameter model for luminosity evolution given by (\ref{6.2}) is adequate for the current subset we analyzed.

\section{Discussion}
\label{sec8}

This paper is concerned with linear regression analysis when the response variable is subject to double truncation. Truncated data can be found in many applications, including those from biomedical researches, economics and astronomy. Most statistical methods for dealing with truncated data are for observations with left or right truncation. The left (right) truncation is relatively easy to handle due to the simple form of re-distribution-to-left (right) algorithm and applicability of counting process-martingale formulation. However, for the doubly truncated data, less technical tools are available, resulting much fewer results.

We propose a novel method to estimate the regression parameter in the linear regression model with doubly truncated responses. To eliminate the bias introduced by double truncation, we extend the Mann-Whitney type loss function for estimating regression parameters by symmetrization. The proposed estimator is obtained by minimizing the extended Mann-Whitney type loss function. The minimization can be done by some standard software packages directly, or by an iterative algorithm with an $L_1$-type minimization in each iteration. The proposed estimator is proved to be consistent and asymptotically normal under some regularity conditions. A simple random perturbation approach is used to get the variance estimator. We also provide a weighted estimation procedure for improving the estimation efficiency. Simulation studies show that the proposed approach works well for moderate sample sizes. The application to the quasar data gives new insights.

In addition to handling multiple covariates, another major advantage of the proposed loss function-based approach to estimation over the test score-based approach of Efron and Petrosian (1999) is that it can easily incorporate a penalty function, such as LASSO, to do variable selection. Note that when LASSO penalty is used, our iterative algorithm is preferable since in each iteration the optimization can still be formulated into an $L_1$-minimization problem, facilitating the computation. It is also of interest to consider if the idea of the proposed approach can be extended to do regression analysis with doubly censored responses, such that discussed by Ren and Gu (1997). These topics certainly warrant future research.

\appendix

\section{Appendix}

\subsection{Two lemmas}

The first lemma is crucial for the intuition towards the proposed loss function $G_n(\beta)$ defined by (\ref{3.4}).

\begin{lem}\label{lem1}
Let $L_i(\beta)$, $R_i(\beta)$ and $e_i(\beta)$, $i=1, \dots, n$ be defined in Section 3. Then the event
\begin{equation}
(L_j-Y_j)\vee(Y_i-R_i)<e_i(\beta)-e_j(\beta)<(R_j-Y_j)\wedge(Y_i-L_i)\label{a.1}
\end{equation}
occurs if and only if $L_j(\beta)<e_i(\beta)<R_j(\beta)$ and $L_i(\beta)<e_j(\beta)<R_i(\beta)$.
\end{lem}

\noindent{\it Proof:} \ We first show ``if''. From $e_i(\beta)<R_j(\beta)$, we have
\begin{equation}
e_i(\beta)-e_j(\beta)<R_j(\beta)-e_j(\beta)=R_j-Y_j.\label{a.2}
\end{equation}
From $L_i(\beta)<e_j(\beta)$, we have
\begin{equation}
e_i(\beta)-e_j(\beta)<e_i(\beta)-L_i(\beta)=Y_i-L_i.\label{a.3}
\end{equation}
Thus, the second inequality of (\ref{a.1}) holds. The second inequality can be shown similarly.

Next we show ``only if''. This can be done by reversing the above argument. From (\ref{a.2}), we obviously have $e_i(\beta)<R_j(\beta)$, while from (\ref{a.3}), we get $L_i(\beta)<e_j(\beta)$. Additionally, from $(L_j-Y_j)\vee(Y_i-R_i)<e_i(\beta)-e_j(\beta)$, we get $e_i(\beta)>L_j$ and $e_j(\beta)<R_i(\beta)$.

The second lemma shows that the choice of $\tfw_{ij}=\tps_{\tn}(\beta,\te_i(\beta)\wedge\te_j(\beta))$ makes the weighted estimation function becomes the log-rank estimation function.

\begin{lem}\label{lem2}
When $\tfw_{ij}=\tps_{\tn}(\beta,\te_i(\beta)\wedge\te_j(\beta))$, where $\tps_{\tn}(b,t)=(\sum_{i=1}^{\tn}I\{\te_i(b)\geqslant t\})^{-1}$, $\tu_{\tn,\tfw}(\beta)$ becomes the log-rank estimating function for $\beta$.
\end{lem}

\noindent{\it Proof:} \ When $\tfw_{ij}=\tps(\te_i(\beta)\wedge\te_j(\beta))$, it can be seen that
\begin{eqnarray*}
\tu_{\tn,\tfw}(\beta)&=&\sum_{i=1}^{\tn}\sum_{j=1}^{\tn}\left(\sum_{k=1}^{\tn}I\{\te_k(\beta)\geqslant \te_i(\beta)\wedge\te_j(\beta)\}\right)^{-1}(\tx_i-\tx_j)\mbox{sgn}\left\{\te_i(\beta)-\te_j(\beta)\right\}\\
&=&-2\sum_{i=1}^{\tn}\sum_{j=1}^{\tn}\left(\sum_{k=1}^{\tn}I\{\te_k(\beta)\geqslant\te_i(\beta)\}\right)^{-1}(\tx_i-\tx_j)I\left\{\te_j(\beta)\geqslant\te_i(\beta)\right\}\\
&=&-2\sum_{i=1}^{\tn}\left(\frac{\sum_{j=1}^{\tn}\tx_iI\left\{\te_j(\beta)\geqslant\te_i(\beta)\right\}}{\sum_{k=1}^{\tn}I\{\te_k(\beta)\geqslant\te_i(\beta)\}}
-\frac{\sum_{j=1}^{\tn}\tx_jI\left\{\te_j(\beta)\geqslant\te_i(\beta)\right\}}{\sum_{k=1}^{\tn}I\{\te_k(\beta)\geqslant\te_i(\beta)\}}\right)\\
&=&-2\sum_{i=1}^{\tn}\left(\tx_i-\frac{\sum_{j=1}^{\tn}\tx_jI\left\{\te_j(\beta)\geqslant\te_i(\beta)\right\}}{\sum_{k=1}^{\tn}I\{\te_k(\beta)\geqslant\te_i(\beta)\}}\right).
\end{eqnarray*}
This completes the proof.

\subsection{Proof of Theorem \ref{thm1}}

We first prove consistency. Let $\bg_n=[n(n-1)]^{-1} G_n$. By the uniform law of large numbers for U-process (Arcones and Gin\'e, 1993), we have that $\bg_n(\beta)$ converges uniformly to $\bg(\beta)$ for $\beta$ over $B$. Since by assumption A3 $\bg(\beta)$ has a unique minimizer $\beta_0$, $\hbeta_n$ must converge to $\beta_0$ as $\bg(\beta)$ is obviously continuous.

The proof of asymptotic normality follows closely the technical developments given in Sherman (1993) for the maximum rank correlation estimator which is also defined as the optimizer of a U-type objective function. In fact, the situation there is more complicated as it deals with a discontinuous objective function. An essential ingredient of Sherman's approach is the quadratic approximation to the objective function.

Following Sherman (1993), define $\tau (z, \beta)=E\xi (Z_i, z; \beta)$. Let $\dot \tau (z, \beta)$ and $\ddot \tau (z, \beta)$ be its first and second derivatives with respect to $\beta$. Then it
can be seen from conditions A1 and A2 that we have
\begin{equation*}
E [\|\dot \tau (Z_i, \beta)\|^2+\|\ddot \tau (Z_i, \beta)\|]<\infty
\end{equation*}
and there exists $K(z)\ge 0$ such that $EK(Z_i)<\infty$ and
\begin{equation*}
\|\ddot \tau (z, \beta)-\ddot \tau (z, \beta_0)\|\le K(z)\|\beta-\beta_0\|.
\end{equation*}
From these and conditions A1-A4, we can verify the four assumptions in Sherman (1993, Theorem 4) from which the asymptotic normality
of $\hbeta_n$ follows.

\subsection{Proof of Theorem \ref{thm2}}

Because of scale invariance for $\hbeta^\ast$ to change in $W_i$, we may assume, without loss of generality, that $\mathsf{E}(W_i)=1/2$. Similarly to the proof of consistency of $\hbeta_n$, we can argue in the same way that $\hbeta^\ast$ is consistent. Let
\begin{eqnarray*}
U_n^\ast(\beta)&=&\sdn\left(W_i+W_j\right) I\left\{L_j(\beta)<e_i(\beta)<R_j(\beta),L_i(\beta)<e_j(\beta)<R_i(\beta)\right\}\\
& &\times(X_i-X_j)\mbox{sgn}\left\{e_i(\beta)-e_j(\beta)\right\}.
\end{eqnarray*}
It is clear that $U_n^\ast(\beta)$ is the derivative of $G_n^\ast(\beta)$. Thus, by definition, $U_n^\ast(\hbeta^\ast)=0$. By the same argument as that of Jin et al. (2001), we can establish asymptotic linearity and therefore, up to an asymptotically negligible term,
\begin{eqnarray*}
0=U_n^\ast(\hbeta^\ast)\approx U_n^\ast(\hbeta_n)+n^2A(\hbeta^\ast-\hbeta_n),
\end{eqnarray*}
or
\begin{eqnarray*}
\sqrt{n}(\hbeta^\ast-\hbeta_n)\approx -n^{-\frac{3}{2}}A^{-1}U_n^\ast(\hbeta_n).
\end{eqnarray*}
Since $U_n(\hbeta_n)=0$, we have
\begin{eqnarray}\label{a.4}
U_n^\ast(\hbeta_n)&=&\sdn\left(W_i-\frac{1}{2}+W_j-\frac{1}{2}\right) I\left\{L_j(\hbeta_n)<e_i(\hbeta_n)<R_j(\hbeta_n),L_i(\hbeta_n)<e_j(\hbeta_n)<R_i(\hbeta_n)\right\}\nonumber\\
& &\times(X_i-X_j)\mbox{sgn}\left\{e_i(\hbeta_n)-e_j(\hbeta_n)\right\}.
\end{eqnarray}
Each summand on the right-hand side of (\ref{a.4}) clearly has mean 0 conditional on data. Standard asymptotic normality for U-statistics can then be used to show that, conditional on the data, $n^{3/2}U_n^\ast(\hbeta_n)$ to a limiting normal distribution. Simple calculation shows that the conditional covariance matrix of $n^{-3/2}U_n^\ast(\hbeta_n)$ given data converges in probability to $V$. Hence Theorem 2 holds.

\subsection{Proof of Theorem \ref{thm3}}

We know that $\hbeta^\fw_n$ is the solution to the estimating equation $U_{n,\fw}(\beta)=0$. By the asymptotic linearity of $U_{n,\fw}$, we have, ignoring an asymptotically negligible term,
\begin{eqnarray*}
0=U_{n,\fw}(\hbeta^\fw_n)\approx U_{n,\fw}(\beta_0)+n^2A_\fw (\hbeta_n^\fw-\beta_0)
\end{eqnarray*}
or $\sqrt n (\hbeta^\fw_n-\beta_0) \approx -n^{3/2}A_\fw^{-1}U_{n,\fw}(\beta_0)$.
Since $n^{-3/2}U_{n,\fw}(\beta_0)$ converges to $N(0,V_\fw)$ by the asymptotic normality of the U-statistics, we get the desired result.

\subsection{Proof of Theorem \ref{thm4}}

Similarly to (A.5) of Jin et al. (2001), we can show that for each $k$, there exists a $p\times p$ matrix $D_k$ such that
\begin{eqnarray*}
\sqrt{n}(\hbeta^\fw_{(k)}-\beta_0)=-n^{-\frac{3}{2}}D_kA^{-1}U_n(\beta_0)-n^{-\frac{3}{2}}(I-D_k)A_\fw^{-1}U_{n,\fw}(\beta_0)+o_p(1).
\end{eqnarray*}
From this and the joint asymptotic normality of $n^{-3/2}U_n(\beta_0)$ and  $n^{-3/2}U_{n,\fw}(\beta_0))$, we conclude that $\sqrt{n}(\hbeta^\fw_{(k)}-\beta_0)$ is asymptotically normal.

\section{References}
\begin{enumerate}[{[}1{]}]
\item Amemiya, T. (1985), {\it Advanced Econometrics}, Harvard University Press, Cambridge, MA.
\item Arcones, M. A., and Gin\'e, E. (1993), ``Limit Theorems for $U$-Processes,'' {\it The Annals of Probability}, {\bf 21}, 1494-1542.
\item Bhattacharya, P. K., Chernoff, H., and Yang, S. S. (1983),``Nonparametric Estimation of the Slope of a Truncated Regression,'' {\it The Annals of Statistics}, 11, 505-514.
\item Bilker, W., and Wang, M.-C. (1996), ``Generalized Wilcoxon Statistics in Semiparametric Truncation Models,'' {\it Biometrics}, 52, 10-20.
\item Chang, M. N., and Yang, G. L. (1987), ``Strong Consistency of a Nonparametric Estimator of the Survival Function with Doubly Censored Data,''  {\it The Annals of Statistics}, 15, 1536-1547.
\item Efron, B., and Petrosian, V. (1999), ``Nonparametric Methods for Doubly Truncated Data,'' {\it Journal of the American Statistical Association}, 94, 824--834.
\item Greene, W. H. (2012), {\it Econometric Analysis (7th Ed.)}, Prentice Hall, Upper Saddle River, NJ.
\item Gu, M. G., and Zhang, C.-H. (1993), ``Asymptotic Properties of Self-Consistent Estimators Based on Doubly Censored Data,''  {\it The Annals of Statistics}, 21, 611-624.
\item Hajek, J. and Sidak, Z. (1967). {\it Theory of Rank Tests}, Academic Press, New York.
\item Harrington, D.P. and Fleming, T.R. (1982), ``A Class of Rank Test Procedures for Censored Survival Data,'' {\it Biometrika}, 69, 133-143.
\item Jin, Z., Lin, D. Y., Wei, L. J., and Ying, Z. (2003), ``Rank-based inference for the accelerated failure time model,'' {\it Biometrika}, 90, 341--353.
\item Jin, Z., Ying, Z., and Wei, L. J. (2001), ``A Simple Resampling Method by Perturbing the Minimand,'' {\it Biometrika}, 88, 381--390.
\item Keiding, N., and Gill, R. D. (1990), ``Random Truncation Models and Markov Processes,'' {\it The Annals of Statistics}, 18, 582-602.
\item Kim, J.P., Lu, W., Sit, T. and Ying, Z. (2013),
``A unified approach to semiparametric transformation models under generalized biased sampling schemes,'' {\it Journal of the American Statistical Association}, {108}, 217-227.
\item Koenker, R., and Bassett, G. (1978), ``Regression Quantiles,'' {\it Econometrica}, 46, 33-50.
\item Lai, T. L., and Ying, Z. (1991a), ``Estimating a Distribution Function with Truncated and Censored Data,'' {\it The Annals of Statistics}, 19, 417-442.
\item Lai, T. L., and Ying, Z. (1991b), ``Rank Regression Methods for Left-truncated and Right-censored Data,'' {\it The Annals of Statistics}, 19, 531-556.
\item Liu, H., Ning, J., Qin, J. and Shen, Y. (2016), ``Semiparametric Maximum Likelihood Inference for Truncated or Biased-Sampling Data'', {\it Statistica Sinica}, 26, 1087-1115.
\item Lynden-Bell, D. (1971), ``A Method of Allowing for Known Observational Selection in Small Samples Applied to 3CR Quasars,'' {\it Monthly Notices of the Royal Astronomical Society}, 155, 95-118.
\item Pollard, D. (1990), {\it Empirical Processes: Theory and Applications} Reginal Conference Series Probability and Statistics 2. Institute of Mathematical Statistics, Hayward, CA.
\item Prentice, R.L. (1978), ``Linear Rank Tests with Right Censored Data,'' {\it Biometrika}, 65, 167-179.
\item Ren, J.-J., and Gu, M. (1997), ``Regression M-Estimators with Doubly Censored Data,'' {\it The Annals of Statistics}, 25, 2638-2664.
\item Shen, P.-S., (2013), ``Regression Analysis of Interval Censored and Doubly Truncated Data with Linear Transformation Models,'' {\it Computational Statistics}, 28, 581-596.
\item Sherman, R. P. (1993), ``The Limiting Distribution of the Maximum Rank Correlation Estimator,'' {\it Econometrica}, 61, 123-138.
\item Tsai, W.-Y. (1990), ``Testing the Independence of Truncation Time and Failure Time," {\it Biometrika}, 77, 167-177.
\item Tsui, K.-L., Jewell, N. P., and Wu, C. F. J. (1988), ``A Nonparametric Approach to the Truncated Regression Problem,'' {\it Journal of the American Statistical Association}, 83, 785-792.
\item Turnbull, B. W. (1976), ``The Empirical Distribution Function with Arbitrarily Grouped, Censored and Truncated Data,'' {\it Journal of the Royal Statistical Society}, Ser. B, 38, 290--295.
\item Wang, M.-C., Jewell, N. P., and Tsai, W.-Y. (1986), ``Asymptotic Properties Of The Product Limit Estimate Under Random Truncation,'' {\it The Annals of Statistics}, 14, 1597-1605.
\item Weinberg, S. (1972), {\it Gravitation and Cosmology}, New York: Wiley.
\item Woodroofe, M. (1985), ``Estimating a Distribution Function with Truncated Data,'' {\it The Annals of Statistics}, 13, 163-177.
\end{enumerate}
\newpage

\begin{sidewaystable}[ht]
\begin{center}
Table 1. Summarized simulation results for the weighted estimates under covariate-independent truncation.
\medskip
{\small\setlength{\tabcolsep}{1mm}
\begin{tabular}{c cc cc cc c cc c ccc c ccc c ccc c ccc c ccc}\hline
&&              && Error        && Weight    &&\multicolumn{4}{c}{Naive}&&\multicolumn{4}{c}{Wilcoxon}&&\multicolumn{4}{c}{$\mbox{log-rank}_3$}\\
&& $n$          && Distribution && Parameter &&  BIAS   & SE    & SEE  &   CP $95\%$ && BIAS   & SE     & SEE    & CP $95\%$&&  BIAS   & SE    & SEE  &   CP $95\%$\\ \hline
&& 200          && Normal   && $\beta_1$ &&0.0117   & 0.1148 & 0.1141  & $94.6\%$ &&0.0181   & 0.2065 & 0.2152 & $95.4\%$ &&0.0233   & 0.2113 & 0.2182 &$95.5\%$ \\
&&              &&          && $\beta_2$ &&$-0.3619$& 0.1042 & 0.1013  & $6.8\%$  &&$-0.0068$& 0.1967 & 0.2085 & $94.6\%$ &&$-0.0054$& 0.2045 & 0.2102 &$94.3\%$ \\
&&              && Logistic && $\beta_1$ &&$-0.0003$& 0.1838 & 0.1858  & $94.2\%$ &&$-0.0016$& 0.3225 & 0.3374 & $94.8\%$ &&0.0086   & 0.3324 & 0.3528 &$95.4\%$ \\
&&              &&          && $\beta_2$ &&$-0.3464$& 0.1695 & 0.1627  & $43.7\%$ &&0.0162   & 0.3042 & 0.3047 & $94.7\%$ &&0.0205   & 0.3127 & 0.3140 &$95.0\%$ \\
&&              && EV       && $\beta_1$ &&0.0011   & 0.1363 & 0.1303  & $93.5\%$ &&0.0040   & 0.2687 & 0.2919 & $95.0\%$ &&0.0029   & 0.2268 & 0.2477 &$94.9\%$ \\
&&              &&          && $\beta_2$ &&$-0.3799$& 0.1224 & 0.1149  & $11.2\%$ &&0.0226   & 0.2797 & 0.3708 & $95.0\%$ &&0.0202   & 0.2420 & 0.3213 &$95.9\%$ \\\\
&& 300          && Normal   && $\beta_1$ &&0.0047   & 0.0938 & 0.0939  & $95.1\%$ &&0.0059   & 0.1657 & 0.1750 & $95.3\%$ &&0.0043   & 0.1724 & 0.1761 &$94.1\%$ \\
&&              &&          && $\beta_2$ &&$-0.3603$& 0.0823 & 0.0837  & $0.9\%$  &&0.0059   & 0.1631 & 0.1680 & $95.6\%$ &&0.0046   & 0.1670 & 0.1687 &$95.8\%$ \\
&&              && Logistic && $\beta_1$ &&0.0008   & 0.1503 & 0.1511  & $95.5\%$ &&0.0024   & 0.2610 & 0.2666 & $94.8\%$ &&0.0050   & 0.2674 & 0.2787 &$96.0\%$ \\
&&              &&          && $\beta_2$ &&$-0.3535$& 0.1355 & 0.1326  & $25.4\%$ &&0.0011   & 0.2338 & 0.2382 & $95.3\%$ &&$-0.0016$& 0.2440 & 0.2494 &$95.4\%$ \\
&&              && EV       && $\beta_1$ &&0.0037   & 0.1065 & 0.1062  & $94.8\%$ &&0.0065   & 0.2064 & 0.2170 & $95.5\%$ &&0.0089   & 0.1797 & 0.1833 &$95.4\%$ \\
&&              &&          && $\beta_2$ &&$-0.3822$& 0.0957 & 0.0938  & $1.8\%$  &&0.0081   & 0.2102 & 0.2285 & $95.2\%$ &&0.0055   & 0.1867 & 0.1963 &$95.0\%$ \\\\
&& 400          && Normal   && $\beta_1$ &&0.0058   & 0.0818 & 0.0816  & $94.5\%$ &&0.0110   & 0.1501 & 0.1504 & $95.1\%$ &&0.0133   & 0.1497 & 0.1508 &$95.4\%$ \\
&&              &&          && $\beta_2$ &&$-0.3627$& 0.0750 & 0.0725  & $0.1\%$  &&0.0002   & 0.1416 & 0.1427 & $95.0\%$ &&$-0.0002$& 0.1398 & 0.1439 &$95.2\%$ \\
&&              && Logistic && $\beta_1$ &&0.0031   & 0.1358 & 0.1312  & $93.5\%$ &&0.0034   & 0.2258 & 0.2294 & $94.8\%$ &&$-0.0004$& 0.2329 & 0.2400 &$95.8\%$ \\
&&              &&          && $\beta_2$ &&$-0.3553$& 0.1132 & 0.1151  & $11.8\%$ &&$-0.0033$& 0.2010 & 0.2048 & $94.6\%$ &&0.0000   & 0.2105 & 0.2140 &$94.5\%$ \\
&&              && EV       && $\beta_1$ &&0.0018   & 0.0936 & 0.0921  & $93.7\%$ &&0.0031   & 0.1799 & 0.1821 & $95.2\%$ &&0.0029   & 0.1538 & 0.1542 &$94.3\%$ \\
&&              &&          && $\beta_2$ &&$-0.3805$& 0.0838 & 0.0816  & $0.4\%$  &&0.0082   & 0.1761 & 0.1869 & $95.8\%$ &&0.0054   & 0.1559 & 0.1624 &$96.2\%$ \\\hline
\end{tabular}
{\footnotesize\begin{tablenotes}
\item[1]Naive: naive estimate by ignoring double truncation; \ Wilcoxon: Wilcoxon weight estimate; \ $\mbox{log-rank}_3$: log-rank weight estimate with $k=3$; \ BIAS: average bias of the estimates; \ SE: standard error of the estimates; \ SEE: average of the estimated standard errors; \ CP $95\%$: empirical coverage probabilities of Wald-type confidence intervals with $95\%$ confidence level.
\end{tablenotes}}}
\end{center}
\end{sidewaystable}

\begin{sidewaystable}[ht]
\begin{center}
Table 2. Summarized simulation results for the weighted estimates under covariate-dependent truncation.
\medskip
{\small\setlength{\tabcolsep}{1mm}
\begin{tabular}{c cc cc cc c cc c ccc c ccc c ccc c ccc c ccc}\hline
&&              && Error        && Weight    &&\multicolumn{4}{c}{Naive}&&\multicolumn{4}{c}{Wilcoxon}&&\multicolumn{4}{c}{$\mbox{log-rank}_3$}\\
&& $n$          && Distribution && Parameter &&  BIAS   & SE    & SEE  &   CP $95\%$ && BIAS   & SE     & SEE    & CP $95\%$&&  BIAS   & SE    & SEE  &   CP $95\%$\\ \hline
&& 200          && Normal   && $\beta_1$ &&0.2316   & 0.1139 & 0.1151  & $48.3\%$ &&0.0183   & 0.2024 & 0.2176 & $95.0\%$ &&0.0217   & 0.2035 & 0.2205  & $95.9\%$ \\
&&              &&          && $\beta_2$ &&$-0.2490$& 0.1022 & 0.0991  & $28.6\%$ &&$-0.0042$& 0.1944 & 0.2053 & $95.1\%$ &&$-0.0035$& 0.1993 & 0.2075  & $94.9\%$ \\
&&              && Logistic && $\beta_1$ &&0.2196   & 0.1834 & 0.1861  & $78.2\%$ &&$-0.0070$& 0.3160 & 0.3381 & $95.5\%$ &&0.0016   & 0.3254 & 0.3508  & $96.6\%$ \\
&&              &&          && $\beta_2$ &&$-0.2358$& 0.1649 & 0.1611  & $69.1\%$ &&0.0129   & 0.2976 & 0.3287 & $93.9\%$ &&0.0168   & 0.3061 & 0.3360  & $93.9\%$ \\
&&              && EV       && $\beta_1$ &&0.2292   & 0.1375 & 0.1308  & $56.7\%$ &&$-0.0003$& 0.2694 & 0.3059 & $95.8\%$ &&0.0013   & 0.2305 & 0.2596  & $95.8\%$ \\
&&              &&          && $\beta_2$ &&$-0.2605$& 0.1203 & 0.1126  & $36.2\%$ &&0.0206   & 0.2711 & 0.3403 & $95.5\%$ &&0.0187   & 0.2357 & 0.2893  & $95.4\%$ \\\\
&& 300          && Normal   && $\beta_1$ &&0.2268   & 0.0957 & 0.0948  & $33.4\%$ &&0.0068   & 0.1689 & 0.1755 & $95.7\%$ &&0.0051   & 0.1733 & 0.1767  & $95.6\%$ \\
&&              &&          && $\beta_2$ &&$-0.2461$& 0.0817 & 0.0817  & $14.2\%$ &&0.0049   & 0.1633 & 0.1630 & $94.9\%$ &&0.0052   & 0.1671 & 0.1635  & $94.8\%$ \\
&&              && Logistic && $\beta_1$ &&0.2190   & 0.1503 & 0.1518  & $68.8\%$ &&$-0.0039$& 0.2556 & 0.2661 & $95.3\%$ &&$-0.0005$& 0.2650 & 0.2778  & $95.8\%$ \\
&&              &&          && $\beta_2$ &&$-0.2444$& 0.1346 & 0.1310  & $53.0\%$ &&$-0.0064$& 0.2299 & 0.2353 & $94.6\%$ &&$-0.0100$& 0.2412 & 0.2455  & $95.4\%$ \\
&&              && EV       && $\beta_1$ &&0.2351   & 0.1051 & 0.1068  & $40.6\%$ &&0.0060   & 0.2083 & 0.2194 & $95.9\%$ &&0.0093   & 0.1827 & 0.1869  & $94.9\%$ \\
&&              &&          && $\beta_2$ &&$-0.2444$& 0.0936 & 0.0919  & $16.8\%$ &&0.0045   & 0.2019 & 0.2169 & $95.4\%$ &&0.0014   & 0.1811 & 0.1875  & $95.2\%$ \\\\
&& 400          && Normal   && $\beta_1$ &&0.2296   & 0.0810 & 0.0821  & $20.4\%$ &&0.0150   & 0.1453 & 0.1505 & $96.2\%$ &&0.0169   & 0.1470 & 0.1510  & $96.2\%$ \\
&&              &&          && $\beta_2$ &&$-0.2454$& 0.0712 & 0.0709  & $6.3\%$  &&0.0032   & 0.1342 & 0.1394 & $95.3\%$ &&0.0019   & 0.1325 & 0.1402  & $94.4\%$ \\
&&              && Logistic && $\beta_1$ &&0.2227   & 0.1362 & 0.1316  & $60.1\%$ &&$-0.0023$& 0.2287 & 0.2294 & $95.0\%$ &&$-0.0047$& 0.2358 & 0.2392  & $95.3\%$ \\
&&              &&          && $\beta_2$ &&$-0.2437$& 0.1123 & 0.1140  & $43.6\%$ &&$-0.0059$& 0.1993 & 0.2033 & $95.6\%$ &&$-0.0018$& 0.2094 & 0.2120  & $95.0\%$ \\
&&              && EV       && $\beta_1$ &&0.2321   & 0.0942 & 0.0928  & $28.8\%$ &&0.0021   & 0.1824 & 0.1844 & $95.2\%$ &&0.0014   & 0.1566 & 0.1575  & $95.1\%$ \\
&&              &&          && $\beta_2$ &&$-0.2614$& 0.0811 & 0.0799  & $9.2\%$  &&0.0044   & 0.1678 & 0.1787 & $95.8\%$ &&0.0022   & 0.1509 & 0.1561  & $95.5\%$ \\ \hline
\end{tabular}
{\footnotesize\begin{tablenotes}
\item[1]Naive: naive estimate by ignoring double truncation; \ Wilcoxon: Wilcoxon weight estimate; \ $\mbox{log-rank}_3$: log-rank weight estimate with $k=3$; \ BIAS: average bias of the estimates; \ SE: standard error of the estimates; \ SEE: average of the estimated standard errors; \ CP $95\%$: empirical coverage probabilities of Wald-type confidence intervals with $95\%$ confidence level.
\end{tablenotes}}}
\end{center}
\end{sidewaystable}

\clearpage

\begin{center}
Table 3. Results from the quasar data: estimation for the model with linear and quadratic term.

\medskip
{\setlength{\tabcolsep}{1mm}
\begin{tabular}{c cc cc cc c cc c ccc c ccc c ccc}\hline
&& Parameter &&  EST     && SE      && $p$-value     \\ \hline
&& $\theta_1$ &&7.6776   && 2.6396 && 0.0036         \\
&& $\theta_2$ &&$-3.3173$&& 2.2408 && 0.1388         \\ \hline
\end{tabular}
{\footnotesize\begin{tablenotes}
\item[1]EST: estimate of the parameter; \ SE: estimated standard error; \ $p$-value: asymptotic $p$-value of the significance test for $\mathsf{H}_0: \theta_j=0$ against $\mathsf{H}_a: \theta_j\ne0$, $j=1,2$.
\end{tablenotes}}}
\end{center}

\clearpage

\clearpage

\end{document}